\documentclass[aps,prl,preprint,tightenlines,superscriptaddress,showpacs,byrevtex]{revtex4}
\usepackage{graphicx} % Include figure files
\usepackage{dcolumn}  % Align table columns on decimal point

\graphicspath{{fig/}}

\begin{document}

\preprint{\vbox{ \hbox{ }
                         \hbox{Belle Preprint 2007-46}
                         \hbox{KEK Preprint 2007-55}
                         \hbox{arXiv: 0711.3252}
}}

\vspace*{0.5cm}

\title{Study of $B \to D^{**} \ell \nu$ with full reconstruction tagging}

\affiliation{Budker Institute of Nuclear Physics, Novosibirsk}
\affiliation{Chiba University, Chiba}
\affiliation{University of Cincinnati, Cincinnati, Ohio 45221}
\affiliation{The Graduate University for Advanced Studies, Hayama}
\affiliation{Hanyang University, Seoul}
\affiliation{University of Hawaii, Honolulu, Hawaii 96822}
\affiliation{High Energy Accelerator Research Organization (KEK), Tsukuba}
\affiliation{Institute of High Energy Physics, Chinese Academy of Sciences, Beijing}
\affiliation{Institute of High Energy Physics, Vienna}
\affiliation{Institute of High Energy Physics, Protvino}
\affiliation{Institute for Theoretical and Experimental Physics, Moscow}
\affiliation{J. Stefan Institute, Ljubljana}
\affiliation{Kanagawa University, Yokohama}
\affiliation{Korea University, Seoul}
\affiliation{Kyungpook National University, Taegu}
\affiliation{\'Ecole Polytechnique F\'ed\'erale de Lausanne (EPFL), Lausanne}
\affiliation{Faculty of Mathematics and Physics, University of Ljubljana, Ljubljana}
\affiliation{University of Maribor, Maribor}
\affiliation{University of Melbourne, School of Physics, Victoria 3010}
\affiliation{Nagoya University, Nagoya}
\affiliation{Nara Women's University, Nara}
\affiliation{National Central University, Chung-li}
\affiliation{National United University, Miao Li}
\affiliation{Department of Physics, National Taiwan University, Taipei}
\affiliation{H. Niewodniczanski Institute of Nuclear Physics, Krakow}
\affiliation{Nippon Dental University, Niigata}
\affiliation{Niigata University, Niigata}
\affiliation{University of Nova Gorica, Nova Gorica}
\affiliation{Osaka City University, Osaka}
\affiliation{Osaka University, Osaka}
\affiliation{Panjab University, Chandigarh}
\affiliation{Saga University, Saga}
\affiliation{University of Science and Technology of China, Hefei}
\affiliation{Seoul National University, Seoul}
\affiliation{Sungkyunkwan University, Suwon}
\affiliation{University of Sydney, Sydney, New South Wales}
\affiliation{Toho University, Funabashi}
\affiliation{Tohoku Gakuin University, Tagajo}
\affiliation{Department of Physics, University of Tokyo, Tokyo}
\affiliation{Tokyo Institute of Technology, Tokyo}
\affiliation{Tokyo Metropolitan University, Tokyo}
\affiliation{Tokyo University of Agriculture and Technology, Tokyo}
\affiliation{Virginia Polytechnic Institute and State University, Blacksburg, Virginia 24061}
\affiliation{Yonsei University, Seoul}
  \author{D.~Liventsev}\affiliation{Institute for Theoretical and Experimental Physics, Moscow} % ITEP
  \author{I.~Adachi}\affiliation{High Energy Accelerator Research Organization (KEK), Tsukuba} % KEK
  \author{H.~Aihara}\affiliation{Department of Physics, University of Tokyo, Tokyo} % Tokyo
  \author{K.~Arinstein}\affiliation{Budker Institute of Nuclear Physics, Novosibirsk} % BINP
  \author{T.~Aushev}\affiliation{\'Ecole Polytechnique F\'ed\'erale de Lausanne (EPFL), Lausanne}\affiliation{Institute for Theoretical and Experimental Physics, Moscow} % ITEP
  \author{A.~M.~Bakich}\affiliation{University of Sydney, Sydney, New South Wales} % Sydney
  \author{V.~Balagura}\affiliation{Institute for Theoretical and Experimental Physics, Moscow} % ITEP
  \author{E.~Barberio}\affiliation{University of Melbourne, School of Physics, Victoria 3010} % Melbourne
  \author{A.~Bay}\affiliation{\'Ecole Polytechnique F\'ed\'erale de Lausanne (EPFL), Lausanne} % Lausanne
  \author{U.~Bitenc}\affiliation{J. Stefan Institute, Ljubljana} % Ljubljana
% ...
% BINP
\collaboration{The Belle Collaboration}

\begin{abstract}
We report a study of semileptonic $B$ decays to $P$-wave $D^{**}$
mesons.  Semileptonic decay to a $D^*_2$ meson is observed for the
first time and its product branching ratio is measured to be
$\mathcal{B}(B^+ \to \bar{D}_2^{*0} \ell^+ \nu) \times
\mathcal{B}(\bar{D}_2^{*0} \to D^- \pi^+) = 0.22 \pm
0.03(\mathrm{stat.}) \pm 0.04(\mathrm{syst.})\%$.  The result is
obtained using fully reconstructed $B$ tags from a data sample that
contains $657 \times 10^6 B\bar{B}$ pairs collected at the
$\Upsilon(4S)$ resonance with the Belle detector at the KEKB
asymmetric-energy $e^+ e^-$ collider.
\end{abstract}


\maketitle

%%%% >>>> keep the final version single-spaced
\tighten
{\renewcommand{\thefootnote}{\fnsymbol{footnote}}}
\setcounter{footnote}{0}

%\section{Introduction}

Heavy Quark Effective Theory (HQET) has proven to be very successful
at describing semileptonic decays of $B$-mesons, especially inclusive
transitions; it allows one to extract $|V_{cb}|$ to better than  2\%
accuracy~\cite{vcb}.  However, some difficulties arise when it is
applied to exclusive decays. For example, certain sum rules (in particular,
the Uraltsev sum rule~\cite{uralsum}) imply the strong dominance of
decays to the narrow excited $D$ mesons over those to the wide excited
$D$ mesons, while some experimental data show the opposite
trend~\cite{uraltsev,bigi}.  However, no complete experimental study
of such semileptonic decays to excited $D$-mesons exists, and thus no
direct comparison with theoretical predictions can be performed.  In
this paper we report on a study of $B \to D^{(*)} \pi \ell \nu$ decays
and measure the excited $D$ contributions to the $D^{(*)}\pi$ final
state.

According to HQET there are two doublets of orbitally excited
($P$-wave) charmed mesons ($D^{**}$), differentiated by their light
quark angular momentum $j_q = 1/2$ or $j_q = 3/2$.  Members of the
$j_q = 3/2$ doublet are predicted to decay only via a $D$-wave and be
relatively narrow, while members of the $j_q = 1/2$  doublet are
predicted to decay only via an $S$-wave and be relatively
broad~\cite{rosner}. The $D^{**}$ states with spin-parity and light
quark angular momentum combinations $0^+$($j_q = 1/2$), $1^+$($j_q = 1/2$),
$1^+$($j_q = 3/2$) and $2^+$($j_q = 3/2$) are usually
labelled $D_0^*$, $D'_1$, $D_1$ and $D_2^*$, respectively. The $D^{**}$
states have previously been observed and studied in hadronic
$B$-decays~\cite{kuzmin}.   Semileptonic $B$ decays to narrow $D_1$
and $D_2^*$ mesons have been studied  by a number of
experiments~\cite{alephetc}.  The semileptonic branching fractions of
$B \to D^{(*)}\pi\ell\nu$ decays were recently measured by
Belle~\cite{livent} and BaBar~\cite{babar}.

%\section{Belle detector}

This measurement is based on a data sample that contains 657 million
$B\bar{B}$ pairs, which corresponds to $605 \,\mathrm{fb}^{-1}$,
collected at the $\Upsilon(4S)$ resonance with the Belle detector
operating at the KEKB asymmetric-energy $e^+ e^-$
collider~\cite{KEKB}.  An additional $68\,\mathrm{fb}^{-1}$ data
sample taken at a center-of-mass energy $60\,\mathrm{MeV}$ below the
$\Upsilon(4S)$ resonance is used to study continuum $e^+ e^- \to q
\bar{q} \ (q=u,d,s,c)$ background.  The Belle detector is a
large-solid-angle magnetic spectrometer that consists of a silicon
vertex detector (SVD), a 50-layer central drift chamber (CDC), an
array of aerogel threshold Cherenkov counters (ACC), a barrel-like
arrangement of time-of-flight scintillation counters (TOF), and an
electromagnetic calorimeter comprised of CsI(Tl) crystals (ECL)
located inside a superconducting solenoid coil that provides a 1.5~T
magnetic field.  An iron flux-return located outside the coil is
instrumented to detect $K_L^0$ mesons and to identify muons (KLM). The
detector is described in detail elsewhere~\cite{Belle}. 
Two inner detector configurations were used. A 2.0\,cm beam pipe and a
3-layer silicon vertex detector was used for the first sample of
152 million $B\bar{B}$ pairs, while a 1.5\,cm beam
pipe, a 4-layer silicon detector and a small-cell inner drift chamber
were used to record the remaining 504 million $B\bar{B}$
pairs~\cite{svd2}.

%\section{Method}

To suppress the large combinatorial background expected in the
reconstruction of final states including a neutrino, we use a full
reconstruction tagging method, which has been improved in comparison
with our previous paper~\cite{livent}.  The first $B$ meson (denoted
as $B_\mathrm{sl}$) is reconstructed in the semileptonic mode of
interest, {\em i.e.} as a combination of all final particles
$D^{(*)} \pi \ell$ except for the neutrino. The remainder of the event is
combined into either a $D^{(*)} \mathrm{n}\, \pi^{\pm}$ ($\mathrm{n} \le 6$) 
or $D^{(*)} \rho^-$ combination to form the tagging $B$ meson
(referred to below as $B_\mathrm{tag}$). Semileptonic decays are
identified by a peak around zero in the missing mass squared spectrum,
$M_{\nu}^2=(P_\mathrm{beams} - P_\mathrm{tag} - P_\mathrm{sl})^2$,
where  $P_\mathrm{beams}$ is the total four-momentum of the beams and
$P_\mathrm{tag}$ and $P_\mathrm{sl}$ are the reconstructed
four-momenta of the $B_\mathrm{sl}$ and $B_\mathrm{tag}$,
respectively. This method provides significantly improved resolution
in the missing momentum in comparison with non-tagging methods, thus
allowing background suppression, separation of different decay modes
and precise calculation of the decay kinematics.

%\section{Selection and reconstruction}

Charged tracks are required to originate from the interaction point
(IP).  Charged tracks positively identified as electrons or muons with
$|\vec{P}| >1.0\,\mathrm{GeV}/c$ are used as leptons.  Charged kaons
are identified by combining information on track ionization loss
($dE/dx$), Cherenkov light yields and time-of-flight
information. No pion identification is required.  Photons are
identified as isolated electromagnetic showers with energies greater
than $50\,\mathrm{MeV}$ that are not matched to any charged track.

$K_S^0$ mesons are reconstructed from $\pi^+ \pi^-$ pairs having an
invariant mass within $\pm 30\,\mathrm{MeV}/c^2$ ($\pm 5 \sigma$) of
the nominal $K_S^0$ mass and a well reconstructed vertex displaced
from the IP.  $\pi^0$ mesons are reconstructed from $\gamma$ pairs
having an invariant mass within $\pm 15\,\mathrm{MeV}/c^2$ ($\pm
3\sigma$) of the $\pi^0$ nominal mass. Such combinations are then
fitted with a $\pi^0$ mass constraint to improve the momentum
resolution.  $\rho^+$ mesons are reconstructed from $\pi^+\pi^0$ pairs
having an invariant mass within $\pm 0.3\,\mathrm{GeV}/c^2$ of the
nominal $\rho^+$ mass.

We reconstruct $D^0$ candidates using six decay modes~\cite{CC}: 
$K^- \pi^+$, $K^- \pi^+ \pi^0$, $K^- \pi^+ \pi^+ \pi^-$, 
$K_S^0 \pi^+ \pi^-$, $K^- K^+$ and $K_S^0 \pi^0$, and $D^+$ candidates 
using decays to $K_S^0 \pi^+$, $K_S^0 \pi^+ \pi^+ \pi^-$, 
$K^- \pi^+ \pi^+$ and $K^+ K^- \pi^+$. A $\pm 15\,\mathrm{MeV}/c^2$ 
interval around the appropriate nominal $D$ mass is used for all 
modes except for $D^0 \to K^- \pi^+ \pi^0$, where we use a 
$\pm 25\,\mathrm{MeV}/c^2$ window, corresponding to about 
$\pm 3 \sigma$ in each case. Selected $D$ candidates are then 
subjected to a mass-vertex constrained fit to improve their 
momentum resolution. $D^{*0}$ candidates are reconstructed via 
decays to $D^0 \pi^0$ and $D^0 \gamma$; the latter mode is not 
used for $B_\mathrm{sl}$ reconstruction because  in this case the 
$D^*\pi$ invariant mass spectrum has a large background. $D^{*+}$ 
candidates are reconstructed in two modes: $D^0 \pi^+$ and $D^+ \pi^0$. 
The mass difference $M(D^*) - M(D)$ for $D^*$ candidates  is required 
to be within $\pm 3\,\mathrm{MeV}/c^2$ (for $D \pi$) and
$\pm 10\,\mathrm{MeV}/c^2$ (for $D^0 \gamma$) intervals around the
$m_{D^*}-m_D$ nominal value (about $3 \sigma$ and $2 \sigma$,
respectively). Reconstructed $D^*$ candidates are subjected to  a
mass-vertex constrained fit.

For the signal $B_\mathrm{sl}$ meson side we form $D\ell$
(normalization mode) and $D^{(*)} \pi \ell$ (signal modes)
combinations.  The energy difference  
$\Delta E \equiv E_\mathrm{tag} - E_\mathrm{CM}$ and 
beam-constrained mass 
$M_\mathrm{bc} \equiv \sqrt{E_\mathrm{CM}^2 - \vec  P_\mathrm{tag}^{2}}$,
(where $E_\mathrm{tag}$ and $\vec{P}_\mathrm{tag}$ are the tag $B$
candidate center-of-mass (CM) energy and momentum and
$E_\mathrm{CM}=\sqrt{s}/2 \simeq 5.29\,\mathrm{GeV}$)  variables are
used for $B_\mathrm{tag}$ selection. The $B_\mathrm{tag}$ signal
region is defined as $M_\mathrm{bc}>5.27\,\mathrm{GeV}/c^2$,  
$|\Delta E|<40\,\mathrm{MeV}$,   which is about $3\sigma$ in both cases. 
The $B_\mathrm{tag}$ candidates  are  subjected to an energy constrained
fit to improve $\vec{P}_\mathrm{tag}$ resolution. We also use events
from the  $50\,\mathrm{MeV}<|\Delta E|<130\,\mathrm{MeV}$ sidebands
for  background subtraction. In these events the $B_\mathrm{tag}$
candidate is fitted with its energy constrained to the center of the
sideband. In the case of  multiple entries in the signal region, the
$B_\mathrm{tag}\,B_\mathrm{sl}$  candidate with the minimum
$\chi^2_\mathrm{tot}$ is chosen, where  $\chi^2_\mathrm{tot}$ is
calculated as a sum over the $\chi^2$ of intermediate  $D$ and $D^*$
mesons mass (mass-vertex) fits and $\chi^2$ of the $B_\mathrm{tag}$
energy fit. The same single candidate selection is applied to
sidebands. The average number of candidates per event is 1.3 in each
case.

\begin{figure}[p]
\includegraphics[width=0.96\textwidth]{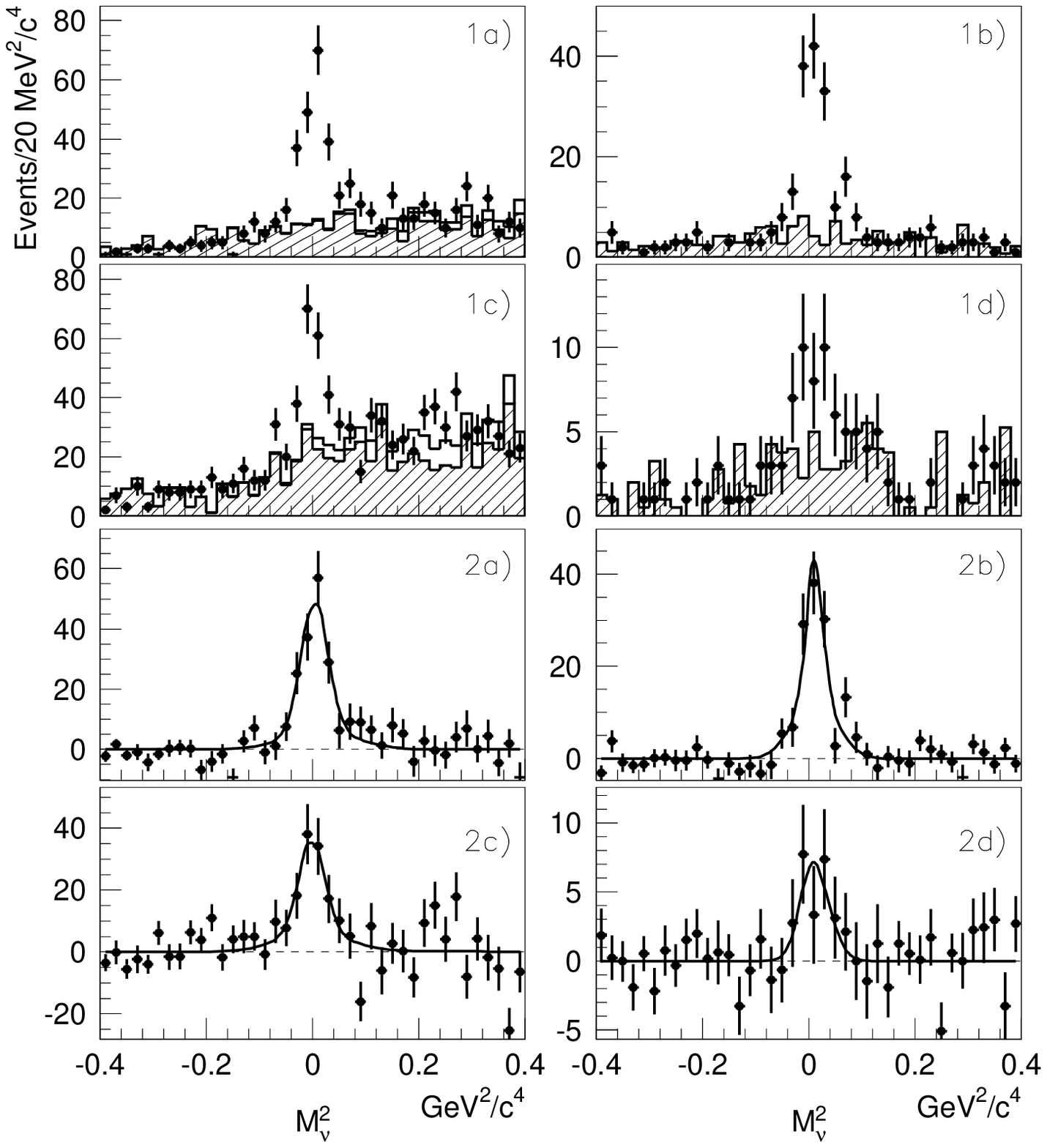}
\caption{$M^2_{\nu}$ spectra before (1) and after (2) background subtraction for:
{\it a)} $B^+ \to D^- \pi^+ \ell^+ \nu$, 
{\it b)} $B^+ \to D^{*-} \pi^+ \ell^+ \nu$, 
{\it c)} $B^0 \to \bar{D}^0 \pi^- \ell^+ \nu$, 
{\it d)} $B^0 \to \bar{D}^{*0} \pi^- \ell^+ \nu$.
The curves are the fits, which are described in the text.}
\label{p:recmass}
\end{figure}

The $M_{\nu}^2$ spectra for the four semileptonic decays $B\to D^{(*)} \pi
\ell \nu$ are shown in Figs.~\ref{p:recmass}, \textit{1a)--1d)} as
points with error bars. Clear peaks are evident in all distributions. 

We divide the backgrounds into the following categories:
\begin{enumerate}
  \item[(1)] Continuum $e^+ e^- \to q \bar{q}$ events.
  \item[(2)] Backgrounds with the $B_\mathrm{tag}$ misreconstructed from 
    particles belonging to the other $B$ meson or fake tracks.
  \item[(3)] $B_\mathrm{sl}$ backgrounds with the $B_\mathrm{tag}$ 
    reconstructed correctly, which can be further separated by their source:
    \begin{enumerate}
      \item[(3a)] Combinatorial background under the $D^{(*)}$ signal  
        from $B_\mathrm{sl}$.
      \item[(3b)] Hadrons misidentified as leptons.
      \item[(3c)] Feed-down from $B \to D^* \pi \ell \nu$ reconstructed 
        as $B \to D \pi \ell \nu$ with lost neutral(s).
    \end{enumerate}
\end{enumerate}

All backgrounds except for (3c) are reliably determined and finally
subtracted directly from the data. Backgrounds (1) and (2) are
estimated using $\Delta E$ sidebands. Continuum data and 
generic $B\bar{B}$ MC simulation show that these backgrounds
have flat $\Delta E$ distributions, thus justifying this
procedure. Background (3a) is subtracted using $D^{(*)}$ ($M(D^{(*)})$)
sidebands. However the region where backgrounds (2) and (3a) 
overlap is subtracted twice by using $\Delta E$ and $M(D^{(*)})$
sidebands. To account for this over-subtraction we use two-dimensional
$\Delta E$ and $M(D^{(*)})$ sidebands. The sum of $M_{\nu}^2$
distributions from one-dimensional $M(D^{(*)})$ and $\Delta E$
sidebands after subtraction of the two-dimensional sidebands are 
illustrated in Figs.~\ref{p:recmass}, \textit{1a)--1d)} by the 
hatched histograms. These histograms represent the sum of 
backgrounds (1), (2) and (3a).

Background (3b) is studied with data by using combinations of
$D^{(*)} \pi$ with high momenta hadrons ($h^+$), where the $h^+$ candidate is
selected with a lepton veto requirement. The combinatorial backgrounds
are subtracted from the observed $M_{\nu}^2$ distributions using
$\Delta E$ and $M(D^{(*)})$ sidebands. The obtained $M_{\nu}^2$
spectra are then multiplied by the known misidentification rate, which
depends on the hadron laboratory momentum (see~\cite{schwanda} for details). 
It is found that the remaining small peak around zero, which is due to 
the contribution of $B \to D^{(*)} \pi \pi^+ (\pi^0)$ decays, is quite 
small ($\sim 0.5-1.0\%$) and we ignore it below, including it as a 
systematic error.

Background (3c) is observed only in the $B \to D \pi \ell \nu$ channels
and is estimated from a MC simulation with normalization fixed to
the data using $B \to D^* (\pi) \ell \nu$ signal yields.
This contribution is plotted in Figs.~\ref{p:recmass}, \textit{1a), 1c)} as 
open histograms.

%\begin{figure}[tbp]
%\includegraphics[width=0.48\textwidth]{fig1.recm.eps}

The background-subtracted $M_{\nu}^2$
distributions are shown in Figs.~\ref{p:recmass}, \textit{2a)--2d)}.
These distributions are fitted with signal functions, the shapes of
which are fixed from MC studies. Fitted signal yields, reconstruction
efficiencies and branching ratios are summarized in
Table~\ref{t:res}. The branching ratios are calculated relative to
the normalization modes $B \to D \ell \nu$ to cancel out the
$B_\mathrm{tag}$ reconstruction efficiency according to the formula:
%$\mathcal{B}(mode) = \mathcal{B}(norm) \times N_\mathrm{mode}/N_\mathrm{norm} \times \epsilon_\mathrm{norm}/\epsilon_\mathrm{mode}$,
\[
\mathcal{B}(mode) = \mathcal{B}(norm) 
\times \frac{N_\mathrm{mode}}{N_\mathrm{norm}} \times \frac{\epsilon_\mathrm{norm}}{\epsilon_\mathrm{mode}},
\]
where $N_\mathrm{norm(mode)}$ and $\epsilon_\mathrm{norm(mode)}$ are
the signal yield and reconstruction efficiency of the normalization
mode (mode of interest) and the normalization mode $\mathcal{B}$ is
taken from the PDG~\cite{pdg}.  Relative efficiencies are obtained
from MC simulation.  Intermediate branching fractions are included, 
while the tagging efficiency is not. The reconstruction and 
background subtraction procedures for the $B \to D \ell \nu$ mode 
are identical to those applied for the studied channels.
The obtained branching fractions are in good agreement with our
previous measurement~\cite{livent} and with BaBar results~\cite{babar}. 
The low efficiency in the last mode is the result of not using the 
$D^{*0} \to D^0 \gamma$ decay channel.

\begin{table}[htb]
\caption{Results for $B \to D^{(*)} \pi \ell \nu$ where
the first error is statistical and the second is systematic.}
\label{t:res}
\begin{tabular}{@{\hspace{0.2cm}}l@{\hspace{0.2cm}}||@{\hspace{0.2cm}}c@{\hspace{0.2cm}}|@{\hspace{0.2cm}}c@{\hspace{0.2cm}}|@{\hspace{0.2cm}}c@{\hspace{0.2cm}}}
\hline \hline
Mode & Yield & Eff.,\% & $\mathcal{B}$(mode),\% \\  
\hline \hline
$B^+ \to \bar{D}^0 \ell^+ \nu$    & $2320 \pm 60$ & 6.4  & $2.15 \pm 0.22$ \footnote[1]{Used as a reference.} \\
$B^+ \to D^- \pi^+ \ell^+ \nu$    & $192 \pm 19$  & 2.8  & $0.40 \pm 0.04 \pm 0.06$ \\
$B^+ \to D^{*-} \pi^+ \ell^+ \nu$ & $123 \pm 14$  & 1.14 & $0.64 \pm 0.08 \pm 0.09$ \\
%\hline
$B^0 \to D^- \ell^+ \nu$                & $760 \pm 30$  & 3.7  & $2.12 \pm 0.20$ \footnotemark[1]\\
$B^0 \to \bar{D}^0 \pi^- \ell^+ \nu$    & $150 \pm 20$  & 3.7  & $0.42 \pm 0.07 \pm 0.06$ \\
$B^0 \to \bar{D}^{*0} \pi^- \ell^+ \nu$ & $22 \pm 8$    & 0.40 & $0.56 \pm 0.21 \pm 0.08$ \\
\hline \hline
\end{tabular}
\end{table}

\begin{figure}[p]
\includegraphics[width=0.96\textwidth]{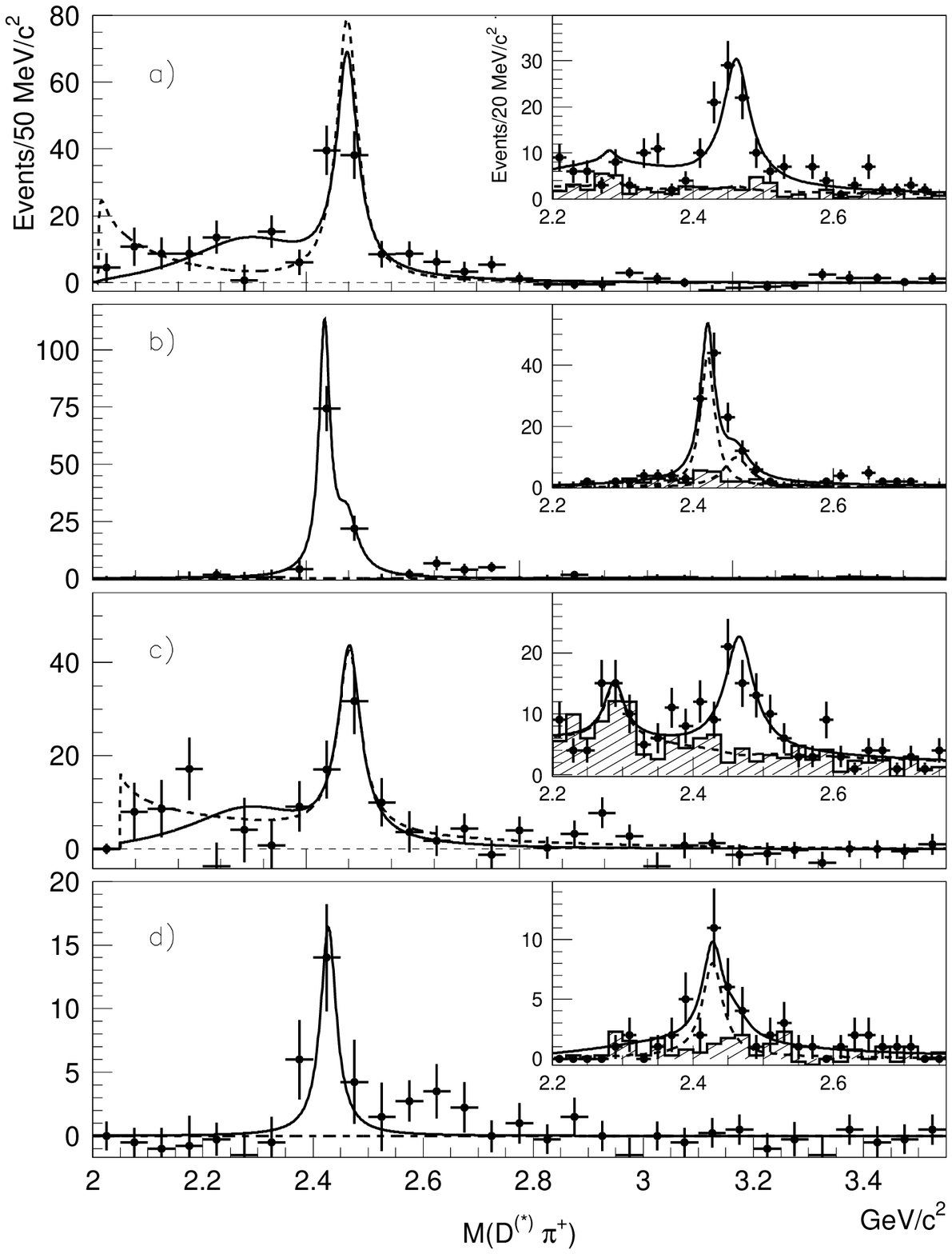}
\caption{Hadronic invariant mass distributions for: {\it a)} $B^+
\to D^- \pi^+ \ell^+ \nu$, {\it b)} $B^+ \to D^{*-} \pi^+ \ell^+ \nu$,
{\it c)} $B^0 \to \bar{D}^0 \pi^- \ell^+ \nu$, {\it d)} $B^0 \to
\bar{D}^{*0} \pi^- \ell^+ \nu$. 
Insets show the distributions before background subtraction in 
the region around the narrow $D^{**}$'s. The background is shown 
as the hatched histogram.
The curves are the fits, which are described in the text.}
\label{p:invmass}
\end{figure}

Signals for semileptonic $B$ decays to orbitally excited $D^{**}$ are
extracted from the $D^{(*)} \pi$ invariant mass distributions. We
define a signal window for $B \to D^{(*)} \pi \ell \nu$ decays by the
requirement $|M^2_{\nu}|<0.1\,\mathrm{GeV}^2/c^4$.  The backgrounds
are estimated in the same way as in the $M^2_{\nu}$ distribution
study. The $D^{(*)} \pi$ invariant mass spectra from the signal window
after subtraction of backgrounds (1-3) are shown in
Fig.~\ref{p:invmass}.  The mass distributions before
background subtraction, restricted to the region near the $j_q=3/2$ states, 
are shown in the insets.

To extract the $D^{**}$ signals we perform simultaneous unbinned
likelihood fits to the signal and background $D^{(*)}\pi$ mass
spectra.  The signal function includes all orbitally excited $D^{**}$
contributing to the given final state ($D_0$ and $D^*_2$ to $D \pi$
and $D_1$, $D'_1$, $D^*_2$ to $D^* \pi$), each of which is described
by a relativistic Breit-Wigner function for a known orbital momenta,
and a non-resonant part described by the Goity-Roberts
model~\cite{goity}. $D^{**}$ masses and widths are fixed to measured
values~\cite{kuzmin}.  To further investigate the $D\pi$
mass spectrum we also test a $D^*_v+D^*_2$ hypothesis.  Despite the
$D^0\pi^+$ mass region corresponding to $D^{*+}$ being excluded from
the study, and while $D^{*0}$ is below the $D^- \pi^+$ threshold, a
virtual $D^*_v$ can be produced off-shell. We describe the $D^*_v$
contribution by a tail of the Breit-Wigner function with floating
normalization. Fit results are shown as a dashed line for this
combination.

A study of the sidebands shows that the background is described by the
sum of a signal function and an exponential.  The resulting signal
function and contributions from the resonances are shown in
Fig.~\ref{p:invmass} as solid and dashed curves, respectively,
superimposed on the background-subtracted mass spectra. In the insets
the solid and dashed curves represent the fitted signal and
background, respectively.  In $B^0 \to D^* \pi \ell \nu$ decays a
small feature may be observed around $2.6\,\mathrm{GeV}/c^2$, which is
absent in $B^+ \to D^* \pi \ell \nu$.  However, the significance of
this feature is small ($2.5\sigma$) and there is no known state there, 
so we do not include a term for it in the fit.  Fitted resonance yields 
and corresponding product branching ratios are listed in
Table~\ref{t:invmass}. The contribution of the non-resonant component
in all cases is consistent with zero.  The $B \to D^{**} \ell \nu$
decay significance is defined as $\sqrt{-2\ln{L_\mathrm{max}}/{L_0}}$,
where $L_0$ is the likelihood value returned by the fit to the
$D^{(*)} \pi$ distribution with the $D^{**}$ contribution fixed to
zero. Our result for $B \to \bar{D}_1 \ell^+ \nu$ is in good agreement
with previous measurements~\cite{alephetc}.  For a
  $D_0^*+D_2^*$ hypothesis the branching ratio of the decay to the
  wide $D^*_0$ is large, in contrast to theoretical
  predictions~\cite{bigi}. However, the present statistics do not
  definitely exclude an interpretation of broadly distributed
  $D \pi^+$ events as the $D^*_v$ tail.

%\begin{figure}[tbp]
%\includegraphics[width=0.48\textwidth]{fig2.invm.eps}

\begin{table}[htb]
\caption{Results of the $D^{(*)} \pi^+$ pair invariant mass study.
$\mathcal{B}(\mathrm{mode}) \equiv \mathcal{B}(B \to D^{**} \ell \nu) \times \mathcal{B}(D^{**} \to D^{(*)} \pi^+)$.
The first error is statistical and the second is systematic.}
\label{t:invmass}
\begin{tabular}
{@{\hspace{0.2cm}}c@{\hspace{0.2cm}}||@{\hspace{0.2cm}}c@{\hspace{0.2cm}}|@{\hspace{0.2cm}}c@{\hspace{0.2cm}}|@{\hspace{0.2cm}}c@{\hspace{0.2cm}}}
\hline \hline
Mode & Yield & $\mathcal{B}$(mode),\% & Signif. \\
\hline \hline
$B^+ \to \bar{D}_0^{*0} \ell^+ \nu$ & $102 \pm 19$ & $0.24 \pm 0.04 \pm 0.06$ & 5.4 \\
$B^+ \to \bar{D}_2^{*0} \ell^+ \nu$ & $94 \pm 13$  & $0.22 \pm 0.03 \pm 0.04$ & 8.0 \\
$B^0 \to D_0^{*-} \ell^+ \nu$       & $61 \pm 22$  & $0.20 \pm 0.07 \pm 0.05$ & 2.6 \\
                                   &              & $<0.4$ @ 90\% C.L.      &      \\
$B^0 \to D_2^{*-} \ell^+ \nu$       & $68 \pm 13$  & $0.22 \pm 0.04 \pm 0.04$ & 5.5 \\
\hline
$B^+ \to \bar{D}_1^{'0} \ell^+ \nu$ & $-5 \pm 11$ & $<0.07$ @ 90\% C.L. & \\
$B^+ \to \bar{D}_1^0 \ell^+ \nu$    & $81 \pm 13$ & $0.42 \pm 0.07 \pm 0.07$ & 6.7 \\
$B^+ \to \bar{D}_2^{*0} \ell^+ \nu$ & $35 \pm 11$ & $0.18 \pm 0.06 \pm 0.03$ & 3.2 \\
$B^0 \to D_1^{'-} \ell^+ \nu$       & $4 \pm 8$   & $<0.5$ @ 90\% C.L. & \\
$B^0 \to D_1^- \ell^+ \nu$          & $20 \pm 7$  & $0.54 \pm 0.19 \pm 0.09$ & 2.9 \\
                                   &              & $<0.9$ @ 90\% C.L.      &      \\
$B^0 \to D_2^{*-} \ell^+ \nu$       & $1 \pm 6$   & $<0.3$ @ 90\% C.L. & \\
\hline \hline
\end{tabular}
\end{table}

\begin{figure}[p]
\includegraphics[width=0.96\textwidth]{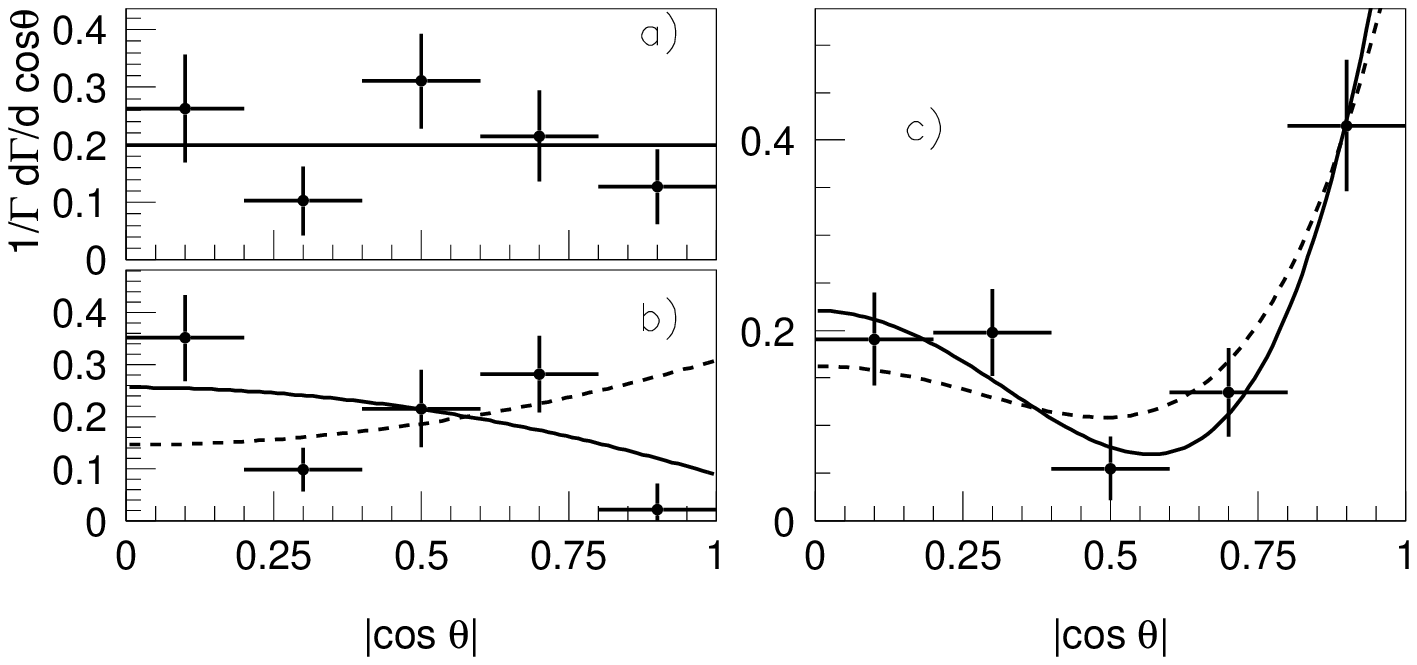}
\caption{Helicity distributions for {\it a)} $D^*_0$, {\it b)} $D^*_v$, 
{\it c)} $D^*_2$.
The curves represent the fits, described in the text.}
\label{p:hel}
\end{figure}

For $D^{*,**}$'s decaying into $D \pi$ we perform a study of the
helicity angle distributions, which is the angle between $\pi$
momentum and the direction opposite to $B_\mathrm{sl}$-momentum in the
$D^{*,**}$ rest frame. To extract the $D^*_v$, $D^*_0$ and the $D^*_2$
helicity distributions we perform a combined fit of the $M(D\pi)$
spectra for $D\pi$ combinations from both $B^+$ and $B^0$ in bins of
helicity angle. The fit procedure is identical to that used for the
$\mathcal{B}(B\to D^{*,**}\ell \nu)$ calculation. The results
corrected for the efficiency are plotted in Fig.~\ref{p:hel}.
$D^*_2$ distributions for $D^*_v$ and $D^*_0$ hypothesis
coincide within errors, so that only that for the $D^*_0+D^*_2$ case
is shown in Fig.~\ref{p:hel}~c.  The $D^*_0$ helicity distribution is
consistent with the $J=0$ hypothesis ($\chi^2/ndf=6.0/4$, where $ndf$
is the number of degrees of freedom). The $D^*_2$
helicity distribution is fitted with the function $a_0^2 |Y^0_2|^2 + 4
a_1^2 |Y^1_2|^2 + 4 a_2^2 |Y^2_2|^2$, where the $Y^i_j$ are spherical
harmonics and $a_0^2 + 4 a_1^2 + 4 a_2^2 =1$. The fit yields $a_0^2 =
0.74 \pm 0.10$, $a_1^2 = 0.04 \pm 0.02$ and $a_2^2 = 0.02 \pm 0.02$;
the fit quality is $\chi^2/ndf=2.0/3$. The fit is consistent with the
assumed quantum numbers and demonstrates that the $D^*_2$ from
semileptonic decay is dominantly in the $s_z=0$ spin
projection. Helicity distributions, predicted by theory, are shown as
dashed lines.  For evaluating the $D^*_v + D^*_2$ hypothesis, the obtained
$D^*_v$ helicity distribution (Fig.~\ref{p:hel}~b) is fitted with the
function $b_0^2 |Y^0_1|^2 + b_1^2 |Y^1_1|^2$. This fit yields $b_0^2 =
0.15 \pm 0.09$, $b_1^2 = 0.85 \pm 0.09$ ($\chi^2/ndf=18.8/4$) in 
poor agreement with expectations from theory, shown as a dashed line.

%\begin{figure}[tbp]
%\includegraphics[width=0.48\textwidth]{fig3.hel.eps}

\begin{figure}[p]
\includegraphics[width=0.96\textwidth]{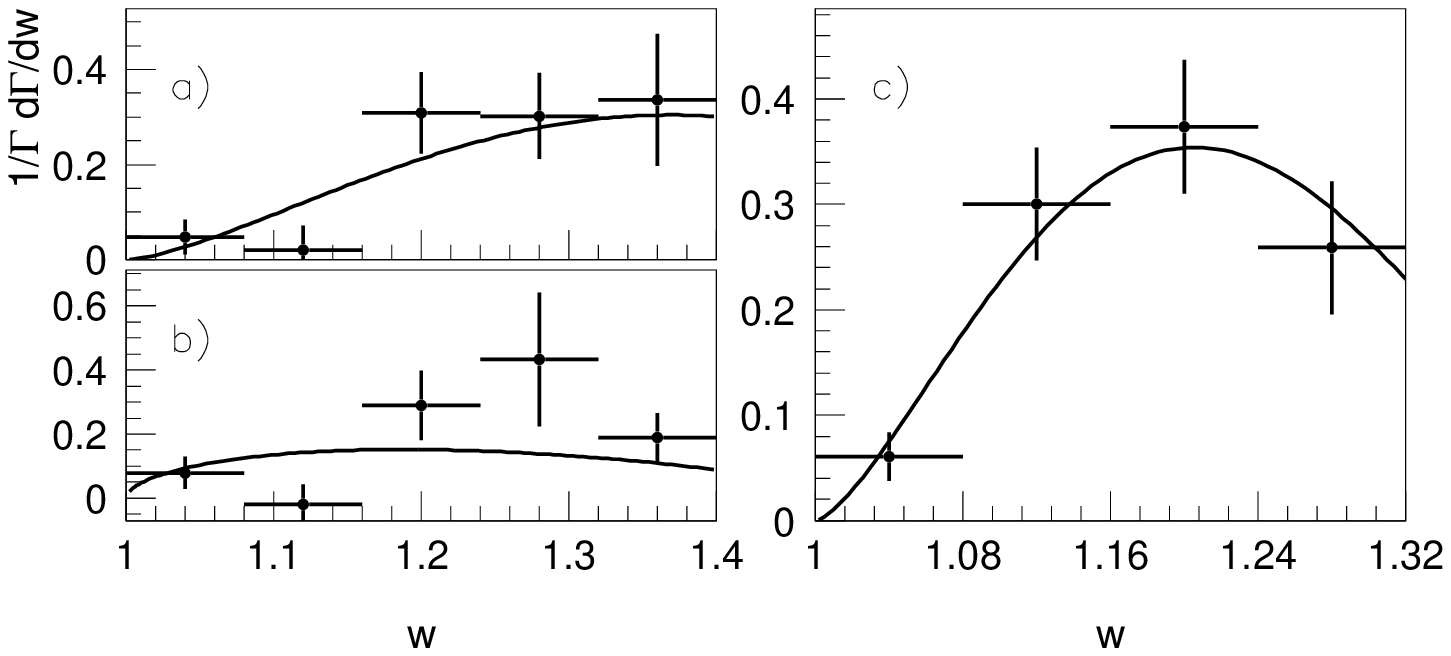}
\caption{$w$ distributions for {\it a)} $D^*_0$, {\it b)} $D^*_v$, {\it c)} $D^*_2$.
The curves are the fits, which are described in the text.}
\label{p:w}
\end{figure}

We also study the dependence of the $B\to D^{**}$ transition on $q^2$
or, equivalently, on the conventional HQET variable $w$, which is the
dot-product of $B$ and $D^{**}$ four-velocities: $w = v_B \cdot v_{D^{**}}$. 
The $w$-dependence is obtained from fits
of $D \pi$ invariant mass in bins of $w$. The results are presented in
Fig.~\ref{p:w}.
As with the helicity study the $D^*_2$ distribution is shown only for 
  the $D^*_0+D^*_2$ hypothesis in Fig.~\ref{p:w}~c.
  The $w$ distribution is fitted according to the model given 
  in Ref.~\cite{yao}. In HQET, the
  matrix elements between the $B$ and $D$ states to leading order in
  $\Lambda_{\mathrm{QCD}}/m_Q$ are expressed in terms of three universal
  Isgur-Wise functions $\xi(w)$, $\tau_{1/2}(w)$ and $\tau_{3/2}(w)$ for $(D, D^*)$, 
  $(D^*_0, D_1')$ and $(D_1,D^*_2)$ doublets, respectively~\cite{yao}. We assume
  a ``pole'' form for $\xi(w)$: $\xi=(2/(1+w))^{2\rho^2}$ and a linear form for 
  $\tau_i(w)$ functions: $\tau_i(w)=\tau_i(1)[1+\hat{\tau}'_i(w-1)]$, and the 
  following relation: $\hat{\tau}_{1/2}'=\hat{\tau}_{3/2}'+0.5$~\cite{veseli}. 
  A simultaneous fit to the $w$-distributions for $D^*_0$ and $D^*_2$ gives 
  $\hat{\tau}_{3/2}'=-1.8 \pm 0.3$. Using the measured branching ratios of 
  $B \to D^*_{0,2}\ell\nu$, we also calculate 
%$\tau_{3/2}(1)$ and $\tau_{1/2}(1)$:
  $\tau_{3/2}(1)=0.75$ and $\tau_{1/2}(1)=1.28$.
  All parameters are in agreement with expectations except for $\tau_{1/2}(1)$,
  which is larger than predicted due to the large value of
  $\mathcal{B}(B \to D^*_0\ell\nu)$.

%\begin{figure}[tbp]
%\includegraphics[width=0.48\textwidth]{fig4.w.eps}

%\section{Systematics}

The systematic error in the calculation of branching fractions due 
to $B_{\mathrm{tag}}$ efficiency uncertainty cancelled out since
normalization  modes were used.
The largest contribution to the systematic error is from uncertainty in 
$D^{**}$ parameters. We perform a $D^{(*)}\pi$ mass study with 
these parameters allowed to float inside their errors to get it.
The $B_{\mathrm{sl}}$ reconstruction
efficiency dependence due to the decay model was studied using two
different signal MC samples generated with the ISGW2~\cite{isgw2} 
and Goity-Roberts~\cite{goity} models.  To estimate the systematic
uncertainty in background subtraction we used two different sets 
of sidebands with appropriate normalizations.  
To estimate interference effects we perform MC study with different 
angle efficiency dependencies.
A summary of the systematic error contributions is  presented in 
Table~\ref{t:sys}.  In total we obtain a 14\% error for  the 
$B \to D^{(*)} \pi \ell \nu$ measurement, a 16\% error for the narrow
$D^{**}$ contribution and 25\% for the measurement of the wide $D^{**}$ 
contribution.

\begin{table}[htb]
\caption{Sources of systematic error.}\label{t:sys}
\begin{center}
\begin{tabular}{l||c}
\hline \hline
Source of error & Systematic error, \%\\
\hline \hline
Reconstruction                 & 2 \\
Model efficiency               & 7 \\
Background subtraction         & 6 \\
Misidentified hadrons          & 1 \\
Interference                   & 7 \\
Normalization branching ratios & 10 \\
$D^{**}$ parameters             & 6--20 \\
\hline
Total                          & 14--25 \\
\hline \hline
\end{tabular}
\end{center}
\end{table}

%\section{Conclusion}

In conclusion, we report measurements of the branching fractions for
$B \to D^{(*)} \pi \ell \nu$ decays. These measurements supersede our
previous results~\cite{livent}.  We also performed an analysis of the
final state $D^{(*)} \pi$ hadronic system and obtained branching
ratios for the $B \to D^{**} \ell \nu$ components. Semileptonic decay
to $D^*_2$ meson is observed and measured for the first time.
Helicity and $w$ distributions are studied for this decay.
We observe a broad enhancement in the $D\pi$ mass
distribution consistent with wide $D^*_0$ production. The branching
ratio of the decay to $B \to D^*_0 \ell \nu$ is found to be large, in
contrast with theoretical predictions~\cite{bigi}.  However there is
no indication of a broad $D'_1$ in the $B \to D^* \pi \ell \nu$
channel, which should be of the same order.  The combined likelihood
of fits to the $D\pi$ mass, helicity and $w$ distributions for
$D_0^*+D_2^*$ hypothesis is higher than that for the $D^*_v+D^*_2$
combination by $2.8\sigma$.

However, the present data sample cannot exclude the interpretation of this
enhancement as a $D^*_v$ tail.

%\section{Acknowledgments}
%----------- Long version, for most papers ----------- 

We would like to thank I.I.~Bigi for useful discussions.

We thank the KEKB group for the excellent operation of the
accelerator, the KEK cryogenics group for the efficient
operation of the solenoid, and the KEK computer group and
the National Institute of Informatics for valuable computing
and Super-SINET network support. We acknowledge support from
the Ministry of Education, Culture, Sports, Science, and
Technology of Japan and the Japan Society for the Promotion
of Science; the Australian Research Council and the
Australian Department of Education, Science and Training;
the National Natural Science Foundation of China under
contract No.~10575109 and 10775142; the Department of
Science and Technology of India; 
the BK21 program of the Ministry of Education of Korea, 
the CHEP SRC program and Basic Research program 
(grant No.~R01-2005-000-10089-0) of the Korea Science and
Engineering Foundation, and the Pure Basic Research Group 
program of the Korea Research Foundation; 
the Polish State Committee for Scientific Research; 
%-> remove for now: under contract No.~2P03B 01324; 
the Ministry of Education and Science of the Russian
Federation and the Russian Federal Agency for Atomic Energy;
the Slovenian Research Agency;  the Swiss
National Science Foundation; the National Science Council
and the Ministry of Education of Taiwan; and the U.S.\
Department of Energy.


\begin{thebibliography}{99}

\bibitem{vcb}
O.L.~Buchmuller, H.U.~Flacher, Phys. Rev. D{\bf 73}, 073008 (2006);
K.~Abe {\it et al.} (Belle Collaboration), BELLE-CONF-0669, hep-ex/0611047;
B.~Aubert {\it et al.} (BaBar Collaboration), arXiv:0707.2670.

\bibitem{uralsum}
N.~Uraltsev, Phys. Lett. B{\bf 501}, 86 (2001).

\bibitem{uraltsev}
N.~Uraltsev, hep-ph/0409125.

\bibitem{bigi}
%I.I.~Bigi {\it et al.}, arXiv:0708.1621.
I.I.~Bigi {\it et al.}, Eur. Phys. J. C{\bf 52}, 975 (2007).

\bibitem{rosner}
J.~Rosner, Comm. Nucl. Part. Phys. {\bf 16}, 109 (1986).

\bibitem{kuzmin}
K.~Abe {\it et al.} (Belle Collaboration), Phys. Rev. D {\bf 69}, 112002 (2004);
A.~Kuzmin {\it et al.} (Belle Collaboration), Phys. Rev. D {\bf 76}, 012006 (2007).

\bibitem{alephetc}
D.~Buskulic {\it et al.} (ALEPH Collaboration), Z. Phys. C {\bf 73}, 601 (1997);
A.~Anastassov {\it et al.} (CLEO Collaboration), Phys. Rev. Lett. {\bf 80}, 4127 (1998);
V.M.~Abazov {\it et al.} (D0 Collaboration), Phys. Rev. Lett. {\bf 95}, 171803 (2005);
J.~Abdallah {\it et al.} (DELPHI Collaboration), Eur. Phys. J. C {\bf 45}, 35 (2006).

\bibitem{livent}
D.~Liventsev {\it et al.} (Belle Collaboration), Phys. Rev. D {\bf 72}, 051109 (2005).

\bibitem{babar}
B.~Aubert {\it et al.} (BaBar Collaboration), arXiv:0708.1738.

\bibitem{KEKB}
S.~Kurokawa and E.~Kikutani, Nucl. Instr. and. Meth. A {\bf 499}, 1 (2003),
and other papers included in this volume.

\bibitem{Belle}
A.~Abashian {\it et al.} (Belle Collaboration), Nucl. Instr. and Meth. A {\bf 479}, 117 (2002).

\bibitem{svd2}
Z.~Natkaniec {\it et al.} (Belle SVD2 Group), Nucl. Instr. and Meth. A {\bf 560}, 1 (2006).

\bibitem{CC}
Throughout this paper, 
the inclusion of the charge-conjugate decay modes is implied
unless otherwise stated.

\bibitem{schwanda}
C.~Schwanda {\it et al.} (The Belle Collaboration), Phys. Rev. D {\bf 75}, 032005 (2007).

\bibitem{pdg}
W.-M.~Yao {\it et al.} (Particle Data Group), J. Phys. G {\bf 33}, 1 (2006).

\bibitem{goity}
J.L.~Goity, W.~Roberts, Phys. Rev. D {\bf 51}, 3459 (1995).

\bibitem{yao}
A.~Le~Yaouanc {\it et al.}, Phys. Lett. B {\bf 520}, 25 (2001).

\bibitem{veseli}
S.~Veseli, M.G.~Olsson, Phys. Lett. B {\bf 367}, 302 (1996). 

\bibitem{isgw2}
D.~Scora and N.~Isgur, Phys. Rev. D {\bf 52}, 2783 (1995);
N.~Isgur, D.~Scora, B.~Grinstein, M.~B.~Wise, Phys. Rev. D {\bf 39}, 799 (1989). 

\end{thebibliography}
\end{document}